\documentclass[12pt,preprint2]{aastex63}

\shorttitle{CO Mapping of Cygnus-X with SMT}
\shortauthors{Baade}
\usepackage{amsmath}
\usepackage{hyperref}
\usepackage{enumitem}

\newcommand{\kms}{km~s$^{-1}$}

\newcommand{\tco}{$^{13}$CO}

\newcommand{\tex}{$T_{\rm ex}$}
\newcommand{\tkin}{$T_{\rm kin}$}

\newcommand{\cmvol}{cm$^{-3}$}

\begin{document}

\title{CO mapping of Cygnus-X -- volume density distribution}

\author[0009-0009-5056-6938]{Jonah C. Baade}
\affil{Steward Observatory, The University of Arizona, Tucson AZ 85721 USA}

\author[0000-0002-8469-2029]{Shuo Kong}
\affil{Steward Observatory, The University of Arizona, Tucson AZ 85721 USA}

\author[0000-0002-6291-7805]{John H. Bieging}
\affil{Steward Observatory, The University of Arizona, Tucson AZ 85721 USA}

\author{Thomas Folkers}
\affil{Steward Observatory, The University of Arizona, Tucson AZ 85721 USA}

\begin{abstract}
We present CO(2-1) and $^{13}$CO(2-1) maps of the Cygnus-X molecular cloud complex using the 10m Heinrich Hertz Submillimeter Telescope (SMT). The maps cover the southern portion of the complex which is strongly impacted by the feedback from the Cygnus OB2 association. Combining CO(1-0) and $^{13}$CO(1-0) maps from the Nobeyama 45m Cygnus-X CO Survey, we carry out a multi-transition molecular line analysis with RADEX and derive the volume density of velocity-coherent gas components. We select those components with a column density in the power-law tail part of the column density probability distribution function (N-PDF) and assemble their volume density into a volume density PDF ($\rho$-PDF). The $\rho$-PDF exhibits a power-law shape in the range of 10$^{4.5}$ \cmvol~$\la n_{\rm H_2} \la$ 10$^{5.5}$ \cmvol~with a fitted slope of $\alpha = -1.12 \pm 0.05$. The slope is shallower than what is predicted by simulations of rotationally supported structures or those undergoing gravitational collapse. Applying the same analysis to synthetic observations with feedback may help identify the cause of the shallow slope. The $\rho$-PDF provides another useful benchmark for testing models of molecular cloud formation and evolution.
\end{abstract}


\section{Introduction}\label{sec:intro}

The density structure of a giant molecular cloud (GMC) is crucial
to our understanding of the star formation in the cloud.
Because of shocks from supersonic turbulence, a GMC typically
contains local over-dense regions that are prone to gravitational
collapse. The collapsing loci will see the birth of the
next-generation young stars. Therefore, a census of the cloud
density, especially those exceeding the critical value for collapse,
is extremely useful for estimating star formation properties,
including, e.g., the star formation efficiency. 

The density probability distribution function
(hereafter $\rho$-PDF) is particularly helpful for the census. 
Due to the multiplicative nature of the density fluctuation, 
the $\rho$-PDF of a GMC with supersonic turbulence is well
approximated by a lognormal distribution
\citep{1994ApJ...423..681V,1997MNRAS.288..145P}
for relatively lower densities. 
The approximation remains valid when considering chemistry
\citep{2010MNRAS.404....2G,2013MNRAS.436.1245F} and magnetic
fields \citep{2008ApJ...682L..97L,2011ApJ...730...40P}.

For higher densities that are gravitationally unstable, 
the $\rho$-PDF is skewed to a power-law shape. 
Using numerical and analytical methods,
\citet[][hereafter K11]{2011ApJ...727L..20K} showed that the power-law
relation originates from the power-law density profile
of spherically symmetric collapsing regions
\citep[also see][]{2014ApJ...781...91G,2020ApJ...903L...2J}.
For a spherically symmetric density
profile $\rho \propto r^{-n}$, the $\rho$-PDF has the 
form of $P(\rho)\propto\rho^{-3/n}$, i.e., a steeper collapsing
density profile gives rise to a shallower power-law $\rho$-PDF.
For instance, a collapsing isothermal sphere with a profile
of $\rho \propto r^{-2}$ \citep[e.g.,][]{1977ApJ...214..488S}
would show a power-law $\rho$-PDF of $P(\rho)\propto\rho^{-1.5}$.
For a collapsing turbulent core with $\rho \sim r^{-1.5}$
\citep{2015ApJ...804...44M}, 
the $\rho$-PDF is $P(\rho)\propto\rho^{-2}$.
K11 also observed a density profile
of $\rho \propto r^{-3}$ for rotationally supported cores
at higher densities ($n_{\rm H_2}>5\times10^9$ cm$^{-3}$)
which indicated a $\rho$-PDF of $P(\rho)\propto\rho^{-1}$. The PDF at the end of the simulation in K11 showed two power-laws,
one being $P(\rho)\propto\rho^{-1.67}$ at intermediate densities,
the other $P(\rho)\propto\rho^{-1}$ at higher densities.
These results were confirmed by the recent simulation work 
from \citet{2021MNRAS.507.4335K}. Including magnetic fields,
\citet{2011ApJ...731...59C} found a $\rho$-PDF power-law slope
of -1.64, which is consistent with the power-law at intermediate
densities in K11.

Observationally, the density structure is usually measured
by the column density (N-PDF) which is directly observable
along each line-of-sight through, e.g., dust emission/absorption
and molecular line emission
\citep[e.g.,][]{2009A&A...508L..35K,2010MNRAS.406.1350F,2013ApJ...766L..17S,2023MNRAS.523.1373M}.
N-PDFs typically exhibit
a lognormal profile at relatively low extinctions with or  
without a power-law tail at higher extinctions. The lognormal 
profile is believed to be a result of supersonic turbulence
\citep[e.g.,][]{1994ApJ...423..681V,2008ApJ...688L..79F,2011MNRAS.416.1436B},
while the power-law tail can result from pressure confinement
\citep[e.g.,][]{1998PhRvE..58.4501P}
and/or gravitational collapse
\citep[e.g.,][]{2011MNRAS.410L...8C,2014MNRAS.445.1575W}.

Recently, \citet[][hereafter S22]{2022A&A...666A.165S} 
measured N-PDF in 29 Galactic regions using Herschel dust 
emission maps. They found that the clouds in general were well fit with a sum of two lognormal profiles at lower
densities and a sum of two power-law tails at higher densities. The massive clouds tend to have a 
shallower second power-law, consistent with the findings
in K11 simulations and \citet{2015MNRAS.453L..41S} observations. 
The shallower power-law is probably due to physical mechanisms
that hinder gas concentration, including, e.g., rotation, 
magnetic fields, and (massive) stellar feedback.

In this paper, we aim to construct the $\rho$-PDF, 
instead of the N-PDF, of the massive star-forming cloud
Cygnus-X. While N-PDF is more straightforward to measure
in observations, $\rho$-PDF is more directly linked to
the underlying physics of the density structure. 
The $\rho$-PDF also provides a pathway to estimating the 
star formation efficiency, especially in active star-forming 
clouds that are dominated by gravity \citep{2014Sci...344..183K}.

To measure the $\rho$-PDF, we simply estimate the gas 
density ($n_{\rm H_2}$) that is responsible for the molecular
line emission through detailed radiative transfer modeling.
As will be introduced in \S\ref{sec:methods}, we use
multi-transition CO and $^{13}$CO lines for the modeling.
The assumption is that the molecular line emission is 
solely excited through the collisional interaction with
the partner H$_2$. The advantage of the method is that the
derived H$_2$ volume density $n_{\rm H_2}$ does not suffer
the depletion problem which happens to molecular column
density measurements \citep[e.g.,][]{2009ApJ...692...91G}, 
as the freeze-out of H$_2$ ice on 
dust grains is rather limited \citep{2017MolAs...9....1W},
if there is any \citep[see discussions in][]{2020ApJ...896L...8S}.
The use of molecular lines also alleviates the problem 
of projection effects in N-PDF studies
\citep[e.g.,][]{2016A&A...587A..74S,2020A&A...641A..53W,2023MNRAS.523.1373M}.

The Cygnus-X molecular cloud complex, at a distance of 
$\sim$1.4 kpc \citep{2012A&A...539A..79R}, is an active star-forming 
region with ample dense gas 
\citep[total mass $\sim10^6$ M$_\odot$,][]{2006A&A...458..855S}
and strong feedback from clusters of O/B stars 
\citep{2008hsf1.book...36R,2015MNRAS.449..741W}.
It provides an unprecedented cosmic lab for the study of
cloud density distribution in the presence of the interaction 
between gravity and massive stellar feedback. As introduced
earlier, a $\rho$-PDF typically contains a power-law profile
at relatively higher densities, which is caused by 
gravitational collapse. However, we lack quantitative 
knowledge about how feedback shapes the $\rho$-PDF. 
Qualitatively, for instance, an expanding H$_{\rm II}$ region
sweeps the vicinity of the driving massive star and creates
compressed, high-density gas, potentially flattening the
power-law. \citet{2015A&A...577L...6S,2015MNRAS.453L..41S}
observed shallower power-law PDF in clouds with stellar 
feedback. In this paper, with the measurement of the $\rho$-PDF
in Cygnus-X, we aim to examine how the strong feedback from the
O-stars (re-)shapes the density distribution. We will focus 
on the power-law part of the $\rho$-PDF at relatively high
densities.

In the following, we first describe our data collection in 
\S\ref{sec:obs}, including our newly-acquired maps of CO(2-1) 
and $^{13}$CO(2-1) using the Heinrich Hertz Submillimeter
Telescope (SMT) and the publicly available Nobeyama 45m Radio Observatory
(hereafter NRO45) maps of CO(1-0) and $^{13}$CO(1-0). Then,
in \S\ref{sec:methods} we describe in detail our multi-line
fitting methods for deriving the gas volume density. Next,
we report our findings in \S\ref{sec:results}. Finally, we
discuss and conclude in \S\ref{sec:discuss} and \S\ref{sec:conclude}, respectively.

\section{Observations and Data Reduction}\label{sec:obs}

\subsection{SMT}\label{subsec:smt}

\begin{figure*}[htbp]
\epsscale{1.15}
\plotone{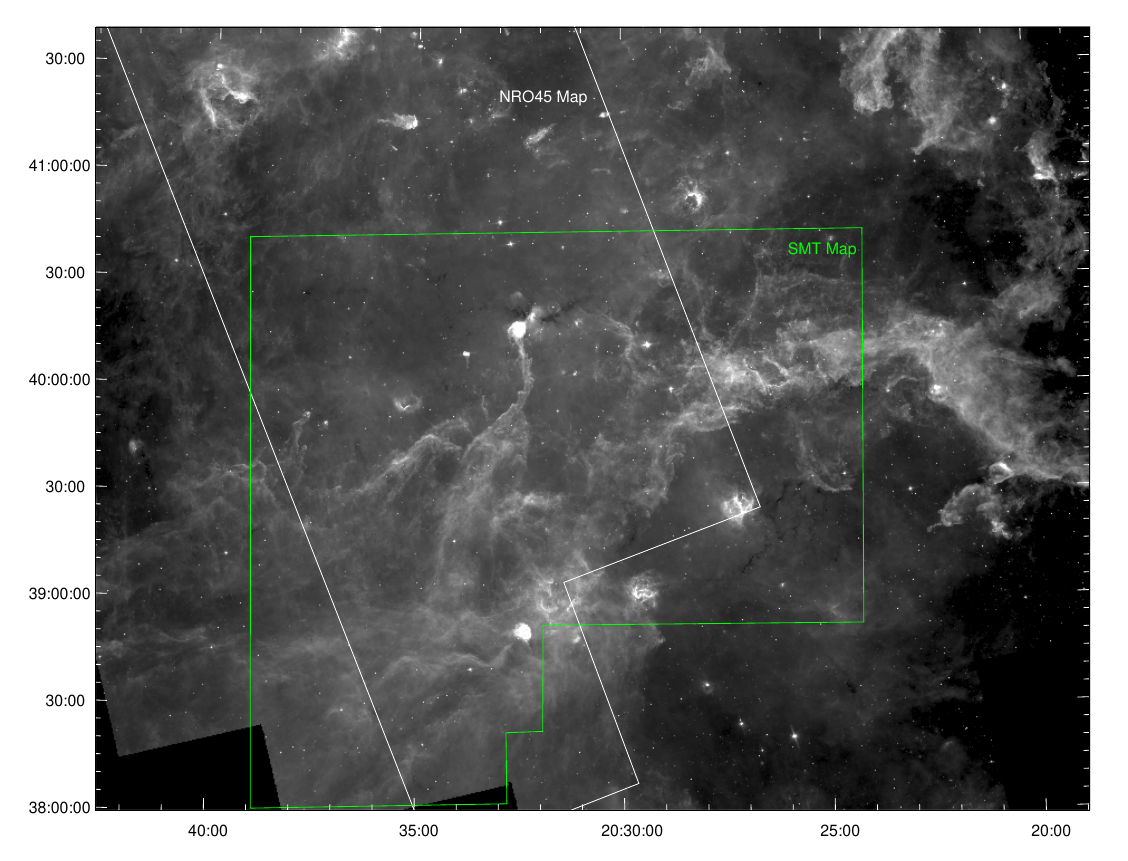}
\vspace{-10pt}
\caption{Spitzer 8 $\mu$m image of Cygnus-X from 
the Spitzer Legacy Survey of the Cygnus-X Region
\citep{2010ApJ...720..679B,2014AJ....148...11K}.
The white polygon shows the NRO45 mapping area 
from the Nobeyama 45m Cygnus-X CO Survey 
\citep{2018ApJS..235....9Y,2019ApJ...883..156T}.
The green polygon shows the SMT mapping area.
Our analysis is done in the overlapping region.}
\label{fig:obsregion}
\end{figure*}

The CO(2-1) and $^{13}$CO(2-1) lines were observed with 
the Heinrich Hertz Submillimeter Telescope (\href{https://aro.as.arizona.edu/?q=facilities/submillimeter-telescope}{SMT})
on Mt. Graham, Arizona, at an elevation of 3200 m. 
The facility is operated by the Arizona Radio Observatories (ARO), 
a division of Steward Observatory at the University of Arizona. The
basic flow of observation and data reduction follows that described in 
\citet{2010ApJS..191..232B,2014ApJS..214....7B,2016ApJS..226...13B}.
In the following, we comment on several key aspects.

We map the southern part of the Cygnus-X complex from November 2019
to June 2021 with a beam size of 32\arcsec.
Figure \ref{fig:obsregion} shows the mapped region
in the green polygon. The region is divided into multiple 
10\arcmin$\times$10\arcmin\ rectangles, each being mapped in 
an on-the-fly (OTF) session for about 1.5 hr. We use the
dual-polarization ALMA Band 6 prototype receiver with the 
spectrometer that simultaneously observes CO(2-1) in 
the upper side-band and $^{13}$CO(2-1) in the lower side-band. 
The data were calibrated and processed as described in 
\citet{2014ApJS..214....7B}. Telescope pointing was checked and
corrected as necessary approximately every few hours.
We use the quarter-MHz filter bank for $^{13}$CO(2-1) which
gives a spectral resolution of 0.34 km s$^{-1}$.
Because of the broad line-width of the CO(2-1) line, 
we use the MHz filter bank for the line to get the baseline fitting.
Typical rms noise values per pixel were 0.09 K in main-beam
brightness temperature per 1.3 km s$^{-1}$ for CO(2-1) and 0.10 K
per 0.34 km s$^{-1}$ for $^{13}$CO(2-1). 

\subsection{Nobeyama 45m data}\label{subsec:nro}

In this paper, we make use of the Nobeyama 45m Cygnus-X survey data
from \citet{2018ApJS..235....9Y,2019ApJ...883..156T}. 
This survey utilized the four-beam receiver FOREST on the 
NRO45 telescope to carry out a multi-line
mapping of the Cygnus-X North and South regions 
(see the white polygon in Figure \ref{fig:obsregion}).
The survey data is publicly available
\href{https://www.nro.nao.ac.jp/~nro45mrt/html/results/data.html}{here}.
For our study, we acquired the cubes for the CO(1-0) and
\tco(1-0) transitions. Out of the several data products,
we selected cubes with a beamsize of
$23\arcsec~$, a spectral resolution of 0.25 \kms~, 
and a velocity range of $-40$ to 40 \kms. 
The choice gives the best match with our SMT data and 
covers the velocity range of Cygnus-X \citep{2006A&A...458..855S}.
The median rms noises for the CO(1-0) and
\tco(1-0) data are 2.37 K and 0.95 K, respectively. 
Using Gaussian kernel methods from the {\tt spectral-cube} and {\tt FITS\_tools} Python packages, we smoothed the NRO45 cubes to 
an effective angular resolution of $32\arcsec~$ and 
a spectral resolution of 0.34 \kms~, to match the 
SMT \tco(2-1) map. We then reprojected them to 
the same spatial and spectral coordinates as the SMT map, 
with a final pixel size of $10\arcsec~$ and 
a velocity range of $-15.9$ to 25.9 \kms~.

\section{Methods}\label{sec:methods}

We use the radiative transfer program 
\href{https://home.strw.leidenuniv.nl/~moldata/radex.html}{RADEX}
\citep{2007A&A...468..627V} to model the observation and 
estimate the gas volume density $n_{\rm H_2}$.
RADEX takes four parameters relevant to our study in its input: 
the kinetic temperature of the gas (\tkin), 
the volume density of the collision partner ($n_{\rm H_2}$),
the molecular column density ($N_{\rm ^{13}CO}$),
and the FWHM line width of the observed velocity component.

For this pilot study, our strategy is to focus on gas parcels
that are in Local Thermodynamic Equilibrium (LTE) so that the volume
density $n_{\rm H_2}$ is the only unknown parameter for RADEX.
Typically, LTE gas tends to be relatively dense ($n_{\rm H_2}\ga10^4$ cm$^{-3}$)
with optically thick CO lines. We analyze the aligned SMT and NRO45 cubes pixel by pixel, fitting $^{13}$CO lines for each. We can approximate the gas kinetic
temperature with the CO line excitation temperature. Then, the $^{13}$CO
column density can be estimated with an LTE equation for the relatively optically thin $^{13}$CO lines.
The $^{13}$CO line width can be obtained through line fitting. With this
information, $n_{\rm H_2}$ becomes the only free parameter, and by varying it we can find a best-fit volume density for the observed emission. The complete procedure is as the following:

\begin{enumerate}[leftmargin=*]
\item We use the \href{http://dendrograms.org/}{\it astrodendro} package \citep{2008ApJ...679.1338R} to find all velocity components in CO(1-0), CO(2-1), $^{13}$CO(1-0), and $^{13}$CO(2-1).
\item Using the {\it astrodendro} results as an initial guess, fit all four lines with Gaussian profiles and get the peak intensity and the velocity dispersion for each velocity component.
\item Cross-match the velocity components between the four lines and keep the components that exist in all of them. Such overlaps between components that emit in all four lines are hereafter referred to as {\it sets} of components.
\item Assuming an optically thick condition, compute the excitation temperature $T_{\rm ex}$ for both CO(1-0) and CO(2-1) and keep the sets whose $T_{\rm ex}$ agrees within 30\%.
\item Compute $^{13}$CO column density N($^{13}$CO) with the CO $T_{\rm ex}$ assuming LTE. We keep the sets whose N($^{13}$CO) are greater than $7 \times 10^{15} \textrm{ cm}^{-2}$ to make sure we are in the power-law regime based on the N-PDF in \citet{2016A&A...587A..74S}. We chose to focus on this power-law regime because our method is less sensitive to low-density gas.
\item With the derived $T_{\rm ex}$, N($^{13}$CO), and $^{13}$CO line dispersion, carry out RADEX modeling \citep{2007A&A...468..627V} of the $^{13}$CO line emission with an initial guess of $n_{\rm H_2}$.
\item Iterate $n_{\rm H_2}$ to best match the $^{13}$CO(1-0) and $^{13}$CO(2-1) line emission between RADEX and observations.
\item Confirm the validity of LTE by checking the $T_{\rm ex}$ between RADEX and observations.
\end{enumerate}
Below, we expand on these steps in more details.

\subsection{Fitting and selecting components}\label{subsec:fit}

\begin{figure}[htbp]
\epsscale{1.15}
\plotone{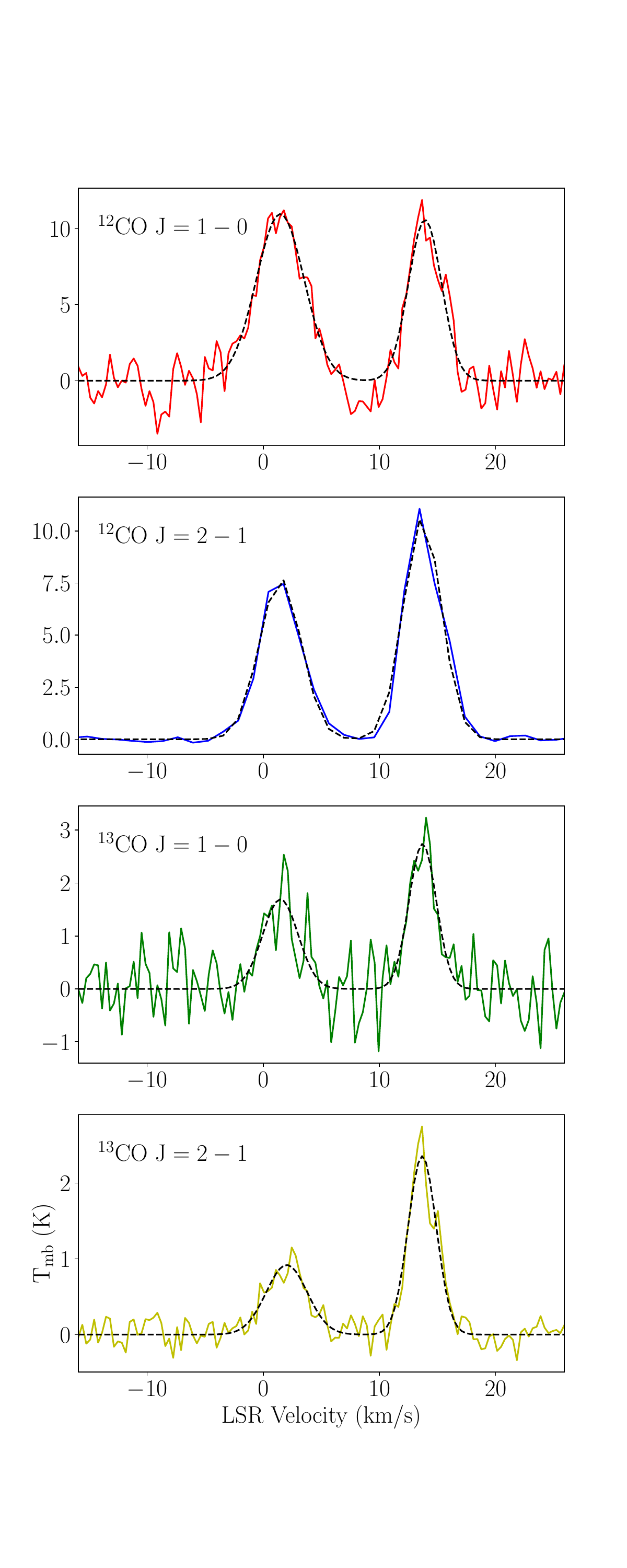}
\vspace{-60pt}
\caption{Plots illustrating Gaussian fit results for the pixel at (RA, DEC) = (20$^{\rm h}$35$^{\rm m}$26$^{\rm s}$, 38\arcdeg27\arcmin31\arcsec). The line data is shown as colored curves while the black dashed line shows the fitted Gaussians. Our program assigned the peaks centered at 2 and 14 \kms~into two sets.}\label{fig:alignedplots}
\end{figure}

In each pixel, spectra in any or all of the four transitions could have multiple peaks centered at different LSR velocities, assumed to correspond to different dense regions along the line of sight. We isolated these peaks using the {\it astrodendro} package (Step 1), using an absolute threshold of $3\sigma$ where $\sigma$ is the pixel-based noise. Each molecular line in each pixel had its own value of $\sigma$, calculated from velocity ranges where the corresponding cubes did not have any apparent bright regions. These ranges were (21, 26) \kms~for CO(1-0); (36, 100) \kms~for CO(2-1); (-16, -11) and (18, 26) \kms~for \tco(1-0); and (-16, -11) and (18, 26) \kms~for \tco(2-1). Pixels with any lines where $\sigma > 2.5$ K, which appeared only around the edges of the data cubes, were discarded. We use a relative threshold of $1.5\sigma$ above the surroundings in velocity space, and a minimum peak width of 3 spectral pixels (2 for the CO(2-1) map, which had a factor of 4 lower spectral resolution) for {\it astrodendro}.

We then used the {\tt curve\_fit} method from the {\tt scipy.optimize} package to fit a sum of Gaussian components to each spectrum (Step 2), using the peaks identified by {\it astrodendro} to provide initial values for the fit and determine the number of peaks to fit. Figure \ref{fig:alignedplots} shows an example of the fitting. If any spectrum out of the four needed more than five components to be adequately fit, we discarded that pixel, under the assumption that the data there was too noisy for a reliable analysis. The peak-fitting approach sometimes fits a low, wide component to the data rather than a small but strong component to match a peak. We excluded these components by discarding any Gaussians that did not reach $3\sigma$ above the pixel's noise level, for consistency with the results from {\it astrodendro}.

Subsequent steps in our data analysis required peak temperatures from all four transitions, so we removed Gaussian components that appeared in only some lines but not in others at the same velocities (Step 3). To do this, we constructed a simple hierarchy in which every Gaussian component in the \tco(2-1) line was matched to the closest component in \tco(1-0), then to CO(2-1) and then CO(1-0). Using the standard velocity deviation $\sigma_v$ of each Gaussian in sequence, we counted pairs of components as aligned if the "later" peak was within 2$\sigma_v$ in velocity of the "earlier". We also excluded all components that shared matches with each other, to avoid situations where a pair of components in one line appeared fully inside a single component in another. After all of these exclusions, we were left with only sets of nested, solitary components at similar velocities, assumed to correspond to a single dense region each. We counted 195,344 sets across 175,592 pixels before further exclusions.

\subsection{Physical conditions}\label{subsec:phys}

\subsubsection{Excitation temperature}

Each set had distinct components in CO(2-1) and CO(1-0). Using their peak temperatures, we estimated two CO excitation temperatures \tex~(Step 4), assuming $\tau \gg 1$, with the following equation \citep[e.g., Eq. 3 in][]{2015ApJ...805...58K}:
\begin{equation}\label{equ:tex}
T_{\rm ex}=\frac{h \nu / k}{\ln \left(1+\frac{h \nu / k}{T_{\rm mb,p}+J_{\nu}(T_{\rm bg})}\right)},
\end{equation}
where $T_{\rm mb,p}$ is the peak main-beam temperature, $T_{\rm bg}$ is the cosmic microwave background temperature of 2.73 K, and $J_{\nu}(T_{\rm bg})=\frac{h \nu / k}{\left(e^{h \nu / k T_{\rm bg}}-1\right)}$ is the Rayleigh-Jeans equivalent temperature of a blackbody of temperature $T_{\rm bg}$.
To limit our modeling to sets that were assumed close to thermal equilibrium and thus amenable to an LTE approximation, we included only sets for which the \tex~ratio of CO(2-1) and CO(1-0) was in the range 0.7-1.3.
We counted 173,188 sets after this restriction.

\subsubsection{$^{13}$CO column density}

We used the \tco(2-1) line to compute the $^{13}$CO column density (Step 5), assuming the LTE condition. The J=2-1 transition has the advantage of not being sensitive to the excitation temperature between 10-40 K \citep[see discussions in][]{2011MNRAS.418.2121G,2016ApJS..226...13B}. We used an LTE equation to estimate $N_{\rm ^{13}CO}$ for each set \citep[e.g., Eqs. 4, 5 in][]{2015ApJ...805...58K}:
\begin{equation}\label{equ:tau}
\tau=-\ln \left(1-\frac{\frac{T_{\rm mb}}{h \nu / k}}{\frac{1}{e^{\frac{h \nu}{k T_{\rm ex}}}-1}-\frac{1}{e^{\frac{h \nu}{k T_{\rm bg}}}-1}}\right)
\end{equation}
\begin{equation}\label{equ:coldens}
\begin{split}
&N=\frac{8 \pi k \nu^{2}}{h c^{3} A_{2 \rightarrow 1}} \frac{e^{\frac{h \nu}{k T_{\rm bg}}}-1}{e^{\frac{h \nu}{k T_{\rm bg}}}-e^{\frac{h \nu}{k T_{\rm ex}}}} \frac{Q}{(2J+1) e^{\frac{-h B_0 J(J+1)}{k T_{\rm ex}}}}\\
&\times \frac{\int \tau(v) d v}{\int\left[1-e^{-\tau(v)}\right] d v} \int T_{\rm mb} d v~({\rm cm}^{-2}),
\end{split}
\end{equation}
where $A_{2 \rightarrow 1}$ is the Einstein coefficient for the J=2-1 transition, and $B_0$ is the rotation constant of $^{13}$CO.
The excitation temperature \tex~used in this calculation was taken from the calculated value for the CO(2-1) transition (if it was close enough to that for the J=1-0 transition in Step 4). The integrals over the main beam temperature $T_{\rm mb}$ and optical depth $\tau(v)$ were solved using the Gaussian fits to the \tco(2-1) spectrum, as integrating over the spectrum itself was problematic when peaks overlapped. The partition function Q is estimated by \citep[e.g., Eq. 6 in][]{2015ApJ...805...58K}
\begin{equation}\label{equ:Q}
Q \equiv \Sigma_{J}(2 J+1) e^{\frac{-h B_{0} J(J+1)}{k T_{\rm ex}}} \simeq \frac{k T_{\rm ex}}{h B_{0}}+\frac{1}{3}
\end{equation}
which is accurate to 10 percent for \tex~$>$~5 K.
See \citet{2015PASP..127..266M} for a comprehensive derivation of the column density calculation.

We restricted our final analysis to sets with a column density $N_{\rm ^{13}CO} \geq 7 \times 10^{15} \textrm{ cm}^{-2}$, equivalent to $N_{\rm H_2} \ga 5 \times 10^{21} \textrm{ cm}^{-2}$ \citep[assuming a H$_2$-to-$^{13}$CO ratio of $\sim7\times10^5$,][]{1999RPPh...62..143W}. These column densities are in accordance with the regime \citet{2016A&A...587A..74S} and S22 identified for a power-law tail in the N-PDF.

\subsubsection{FWHM line width}

Each set had a single Gaussian component in the \tco(2-1) transition. We used this component to calculate a FWHM line width (from Step 2) to enter into RADEX.

\subsubsection{Kinetic temperature}

Under the assumption that the gas in the sets of interest was in LTE, we set \tkin~for the RADEX model equal to the CO(2-1) excitation temperature (from Step 4).

\subsection{RADEX fitting}\label{subsec:radexfit}

We aimed to constrain the $\text{H}_2$ volume density for a set by reproducing the \tco~line intensities (Step 6). In RADEX, the background radiation temperature was set to 2.73 K, and only $\text{H}_2$ was entered as a collision partner.

We configured RADEX with the default escape probabilities for a uniform sphere. As a result, we are essentially estimating the line-of-sight weighted average of the volume density of the gas parcel that contributed to an observed set of peaks. For a parcel with a monotonic density profile, our estimation samples the volume density in a shell of the parcel. Since we are sampling the volume density at different locations across Cygnus-X, the final $\rho$-PDF is considered a reasonable representation of the underlying $\rho$-PDF of the entire cloud at the scale of the beam size (32\arcsec, corresponding to 0.22 pc at a distance of 1.4 kpc).

\begin{figure*}[htbp]
\epsscale{1.}
\plottwo{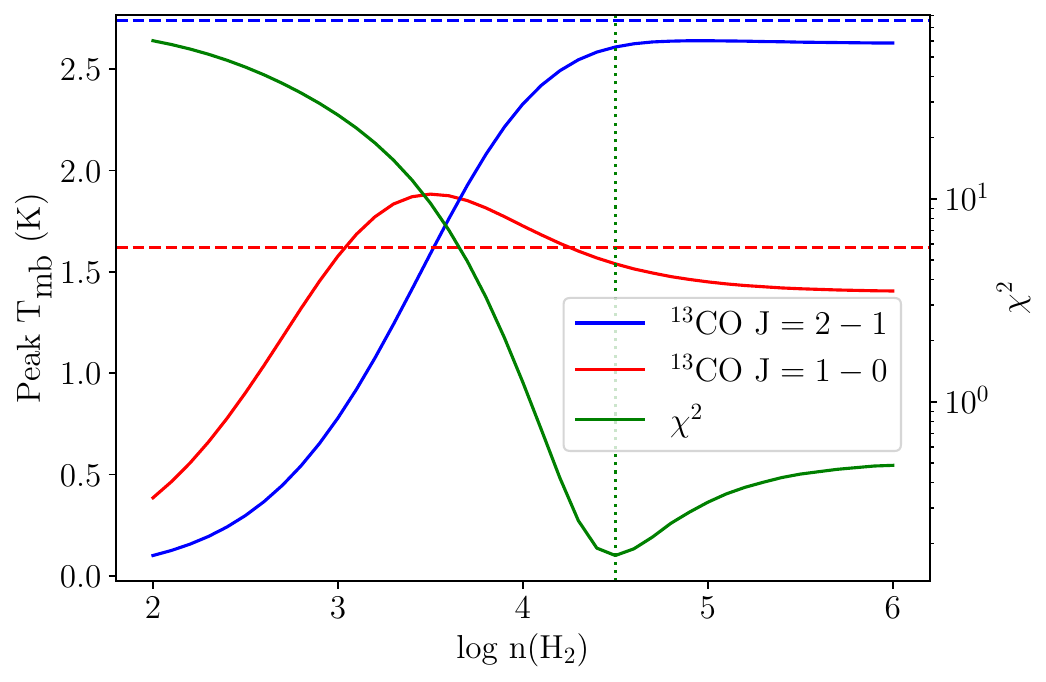}{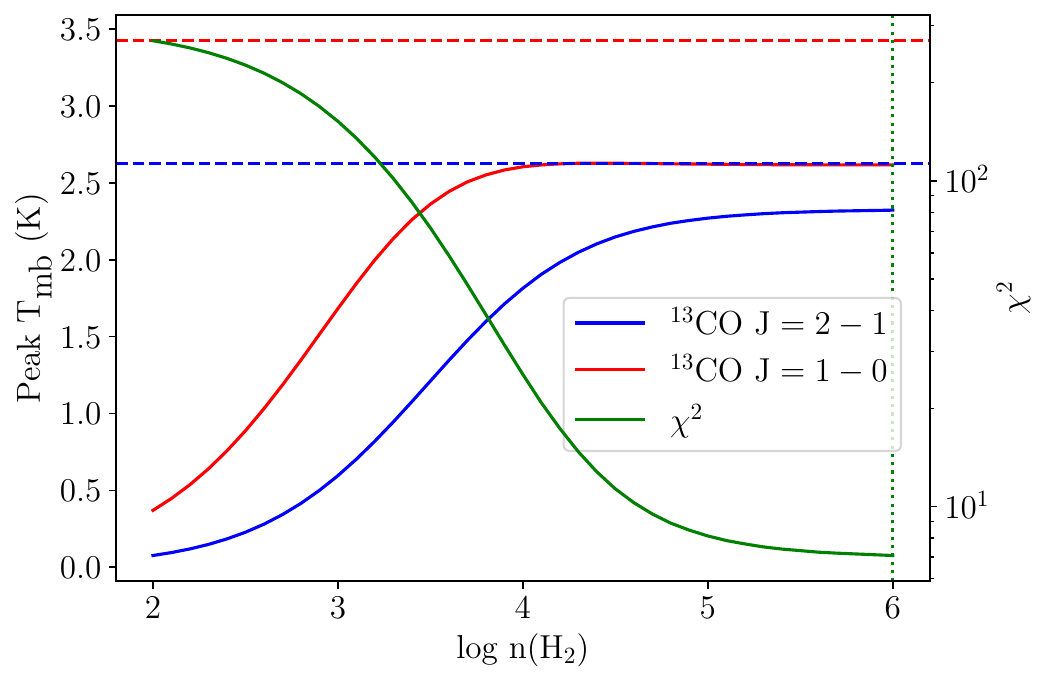}
\caption{Plots illustrating the $n_{\rm H_2}$ fitting process. The left panel is for the set at the pixel at (RA, DEC) = (20$^{\rm h}$37$^{\rm m}$59$^{\rm s}$, 39\arcdeg55\arcmin01\arcsec). The solid red and blue lines indicate the peak $T_{\rm mb}$ that RADEX outputs for each value of $n_{\rm H_2}$, while the dashed lines are at the corresponding peak $T_{\rm mb}$ observed along the line of sight. The dotted green line, at $\textrm{n} = 10^{4.5}~$\cmvol, marks the best fit density for this set, where the $\chi^2$, shown for each density in solid green and labeled on the right, is minimized. The right panel is for the set at the pixel at (RA, DEC) = (20$^{\rm h}$34$^{\rm m}$37$^{\rm s}$, 40\arcdeg30\arcmin21\arcsec). The best fit density from RADEX fitting was $\textrm{n} = 10^{6}~$\cmvol, which was the highest input in our grid.}\label{fig:fitprocess}
\end{figure*}

Because the peak radiation temperatures of \tco(1-0) and \tco(2-1) were known for each set, while $n_{\rm H_2}$ was unknown, we parameterized over H$_2$ volume density and created a grid of identical inputs with only the volume density changed. We tested a range for $n_{\rm H_2}$ of $10^2$-$10^6$ \cmvol, with a logarithmic step size of 0.1 dex. We attributed a best-fit volume density (Figure \ref{fig:fitprocess}) to a set by using the value of $n_{\rm H_2}$ which produced the smallest $\chi^2$ value between the observed and modeled peak radiation temperatures in both \tco\ lines (Step 7). The $\chi^2$ is computed as 
\begin{equation}
    \chi^2=\Sigma \frac{(O-E)^2}{\sigma^2},
\end{equation}
where {\it O} stands for observation and {\it E} stands for model. $\sigma$ is the spectral noise. Currently, we are limited by just two \tco~line transitions. Adding more transitions in the future will give stronger constraints. While the vast majority ${\rm(\sim 88\%)}$ of sets had a local minimum $\chi^2$ within our test range, the remainder showed $\chi^2$ approaching an asymptote as we approached the maximum test volume density of $10^6$ \cmvol. In the final output, these sets were placed in the $10^6$ \cmvol~group.
After fitting, we counted 171,354 sets across 156,371 pixels (with the discrepancy amounting to 1,834 sets that failed the integration for column density). Most of the sets were poorly fit: only 5,480 of them reached $\chi^2 < 10$ at any point (including 1,079 which did not have a local minimum), across 5,467 pixels. However, this subset covers the main spatial extent of the region we have observed, sampling primarily from filaments and dense clouds. This provides reason to believe that the H$_{2}$ density distribution we measure reflects the true distribution throughout the denser parts of Cygnus-X South.

Our choice of cutoff at $\chi^2 < 10$ was arbitrary; it represents the best ${\rm\sim 17\%}$ of fits meeting the column density requirement, but there is no reason to suspect this is the best requirement for closeness of fit to use. We repeated the final step of the procedure for different $\chi^2$ cutoff values, spaced by powers of two from 2.5 to 1,280. Further increasing the $\chi^2$ cutoff value did not increase the number of sets with $N_{\rm ^{13}CO} \geq 7 \times 10^{15} \textrm{ cm}^{-2}$ (required by criterion 5 in \S\ref{sec:methods}). The results are shown in Figure \ref{fig:chisq_counts} and discussed in Section \ref{subsec:pdf}.

\section{Results}\label{sec:results}

\subsection{Global spectra}

\begin{figure*}[htbp]
\epsscale{1.1}
\plotone{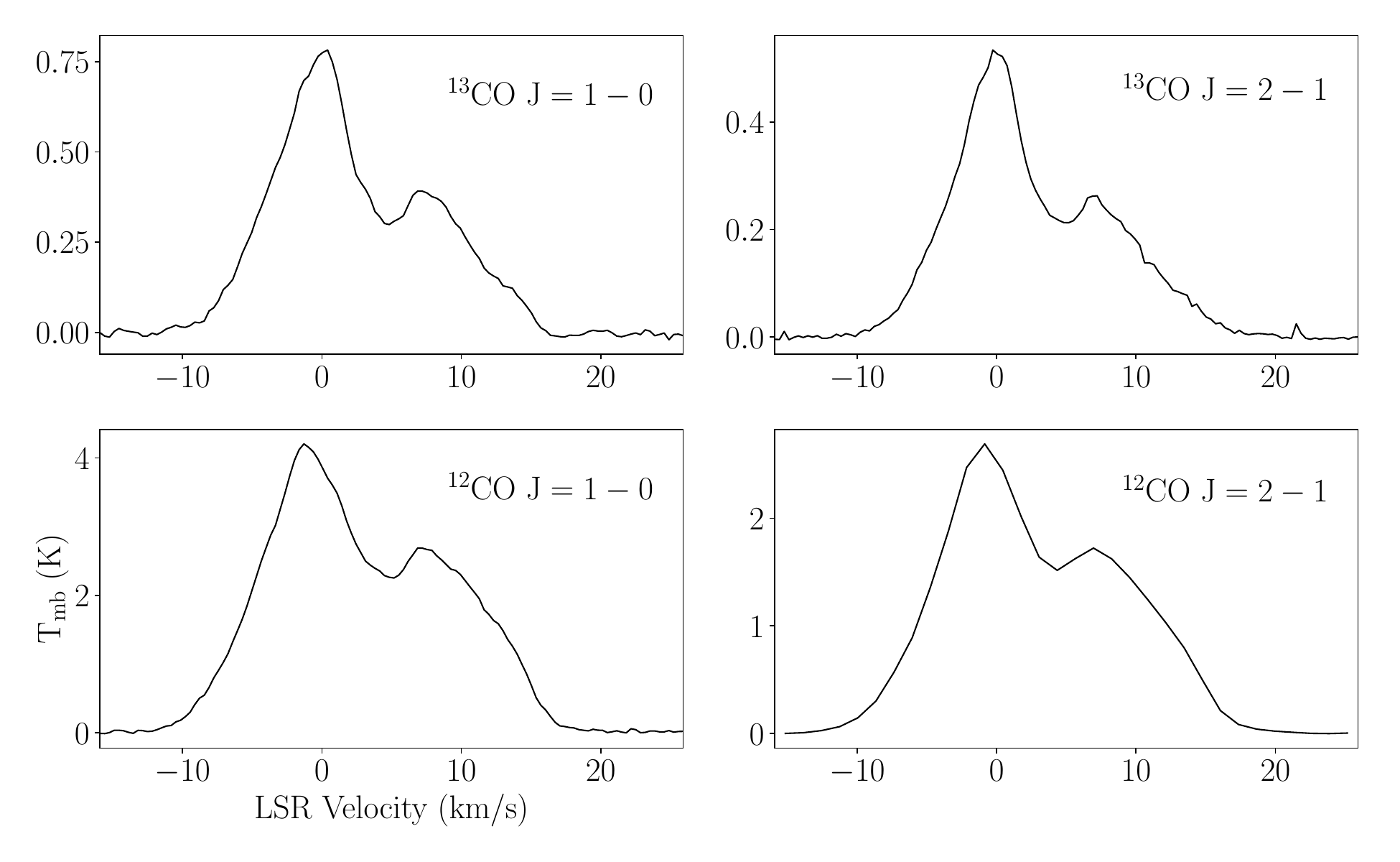}
\caption{Averaged spectra for all four molecular lines across the entire region of mutual overlap between cubes. Pixels with $\sigma > 2.5$ K, which appeared only around the edges of the data cubes, were discarded.}\label{fig:avgspectra}
\end{figure*}

Figure \ref{fig:avgspectra} shows spectra of the four lines we studied, averaged across the area where all four of these lines were present. All four lines have a strong component at or near 0 \kms~ and a weaker component near 7 \kms~, but with significant emission across a much wider band of velocities, generally spanning the range $-10$ to 15 \kms~. Each of the \tco\ lines peaks at about one-fifth the main beam temperature of its counterpart in CO at the same transition.

\subsection{Integrated intensity and first-moment maps}

\begin{figure}
\epsscale{1.16}
\plotone{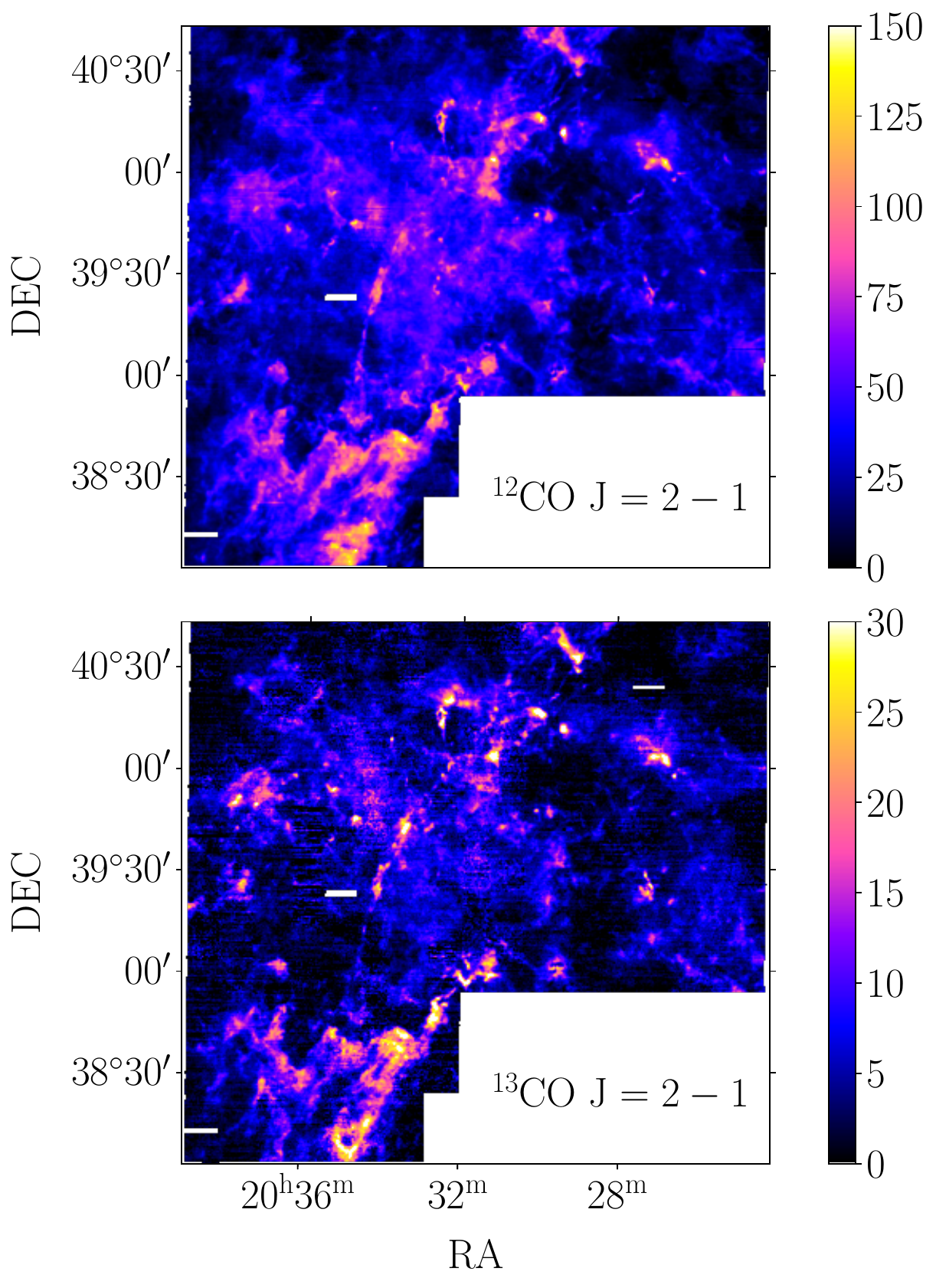}
\caption{Integrated intensity maps for the CO(2-1) and \tco(2-1) data cubes. The integrated velocity range is $-16$ to 26 \kms~. The color bar has units of K \kms~.}\label{fig:integrated_intensities}
\end{figure}

\begin{figure}
\epsscale{1.16}
\plotone{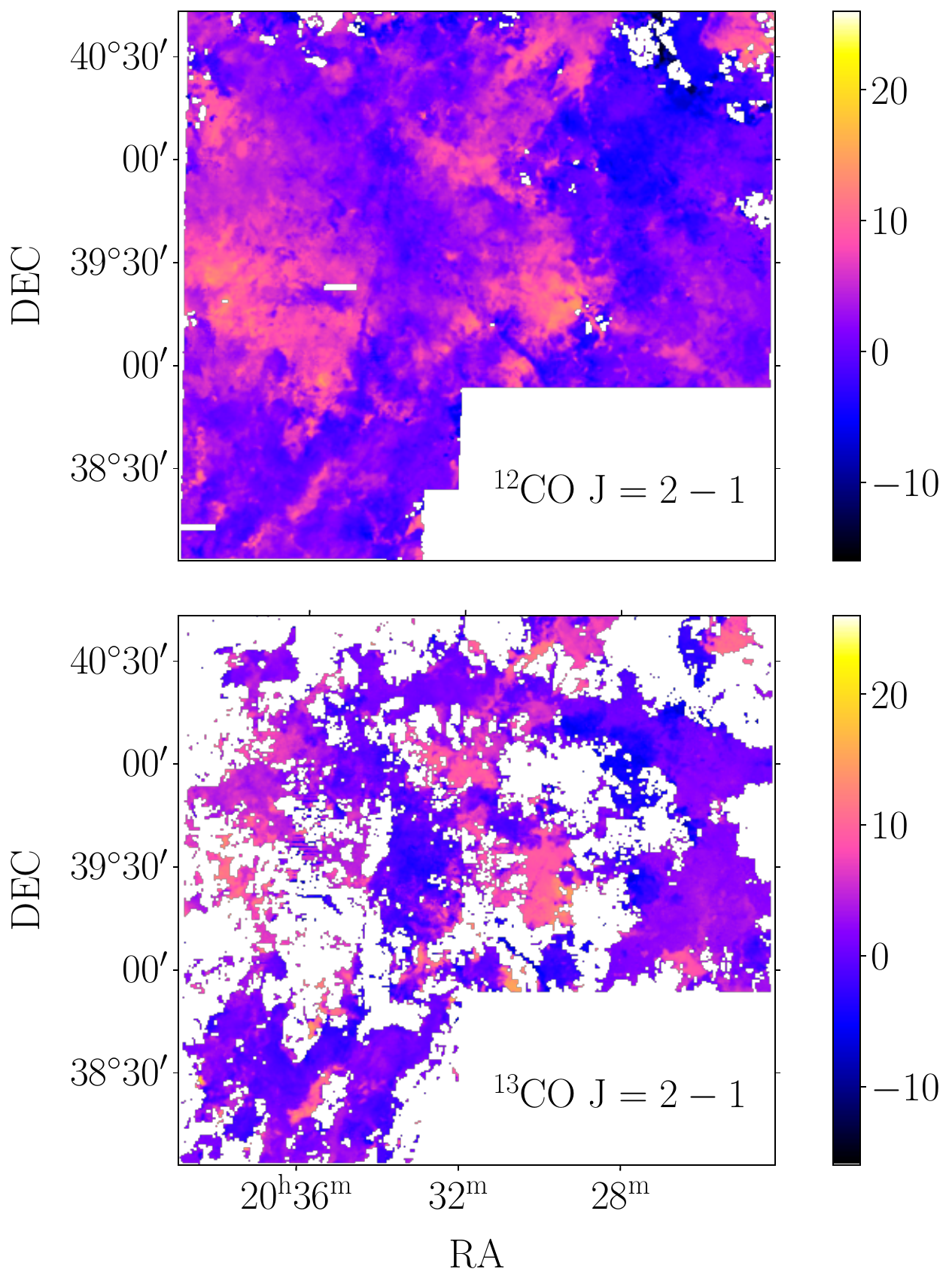}
\caption{First-moment maps for the CO(2-1) and \tco(2-1) data cubes. The integrated velocity range is $-16$ to 26 \kms~. Velocities in pixels with a peak intensity less than $3\sigma$ above the noise level were masked out. The color bar has units of \kms~.}\label{fig:first_moments}
\end{figure}

In Figure \ref{fig:integrated_intensities}, we show maps of the integrated line intensity for CO(2-1) and \tco(2-1), both from SMT observations. In general, the integrated intensity traces the same features in both maps, and it is about five times higher in CO than in \tco\ everywhere.

Figure \ref{fig:first_moments} shows maps of the first moment, the intensity-weighted mean velocity of the gas along the line of sight, for CO(2-1) and \tco(2-1). Both maps show large regions where the mean velocity is zero or slightly negative, and other regions where it is in the vicinity of 10 \kms~, which is consistent with the averaged spectra in Figure \ref{fig:avgspectra}. Some features are visible in both integrated intensity and first-moment maps: in particular, the filament at the bottom of the images is visible as a bright spot in both integrated intensity maps and as a negative-velocity region in the first moment maps. The map in \tco\ is missing many pixels because the emission intensity in those pixels was never more than $3\sigma$ above the noise level. This is a consequence of emission in \tco\ being weaker overall.

\subsection{$\rho$-PDF}\label{subsec:pdf}

\begin{figure}
\epsscale{1.16}
\plotone{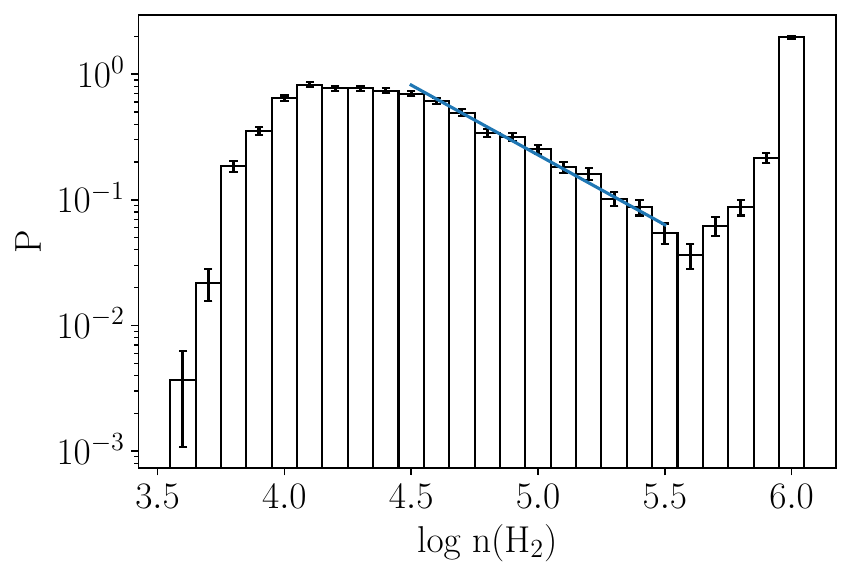}
\caption{$\rho$-PDF derived from best-fit volume densities for each set. Only those with a best-fit $\chi^2 < 10$ were included. The blue line displays a linear-regression power law fit to the data between log($n_{\rm H_2}$) = 4.5 and 5.5.}\label{fig:voldens}
\end{figure}

In Figure \ref{fig:voldens}, we show the $\rho$-PDF of the best-fit volume densities for the 5,480 sets that reached $\chi^2 < 10$. The plot, in logarithmic scales, covers a range in log($n_{\rm H_2}$) from 3.6--6.0, in logarithmic increments of 0.1 dex; no well-fit sets had best-fit volume densities less than $10^{3.6}$ \cmvol. A large number of sets whose $\chi^2$ values did not reach a local minimum within our test range, and were therefore placed in the $10^{6}~$\cmvol\ group, can be attributed to saturation while performing the fitting with RADEX. For example, in Figure \ref{fig:fitprocess} right panel, the calculated peak $T_{\rm mb}$ for each transition gradually approached the observed peak $T_{\rm mb}$, but stopped rising before they could match. This could indicate that the density is actually over $10^{6}~$\cmvol\ in that area. For instance, \citet{2016A&A...587A..74S} detected high-density tracers in the cloud. In our case, a higher transition line would break the degeneracy.

The value of the probability distribution function P at each bin was calculated from the ratio of the number of sets in it divided by the total (in this case 5,480), and then divided by the bin width of 0.1 accordingly to make $\int_{3.6}^{6.0} P~\mathrm{d}\log(n_{\rm H_2}) = 1$. We calculated a line of best fit with linear regression, using the binned volume densities as the x-coordinate in $\log(n_{\rm H_2})$-space, and the bin heights as the y-coordinate in $\log(P)$-space. This is consistent with the corresponding column-density log-log fit between $A_V \propto N_{\rm H_2}$ and $P$ for power-law tails seen in e.g. S22.

At low densities, the distribution appears lognormal, but the shape is more likely a result of a lack of detected sets of low enough intensity to correspond to these densities. At high densities, the distribution appears to be a straight line in the log-log plot, indicating a power-law relation in the $\rho$-PDF ($P \propto \rho^{\alpha}$). A linear-regression fit between log($n_{\rm H_2}$) = 4.5 and 5.5 gives a power-law slope of $\alpha = -1.122 \pm 0.048$. Above these densities, the frequency rises again until our grid's maximum volume density of $10^{6.0}$ \cmvol, where the greatest number of sets matched. 

\begin{figure}
\epsscale{1.16}
\plotone{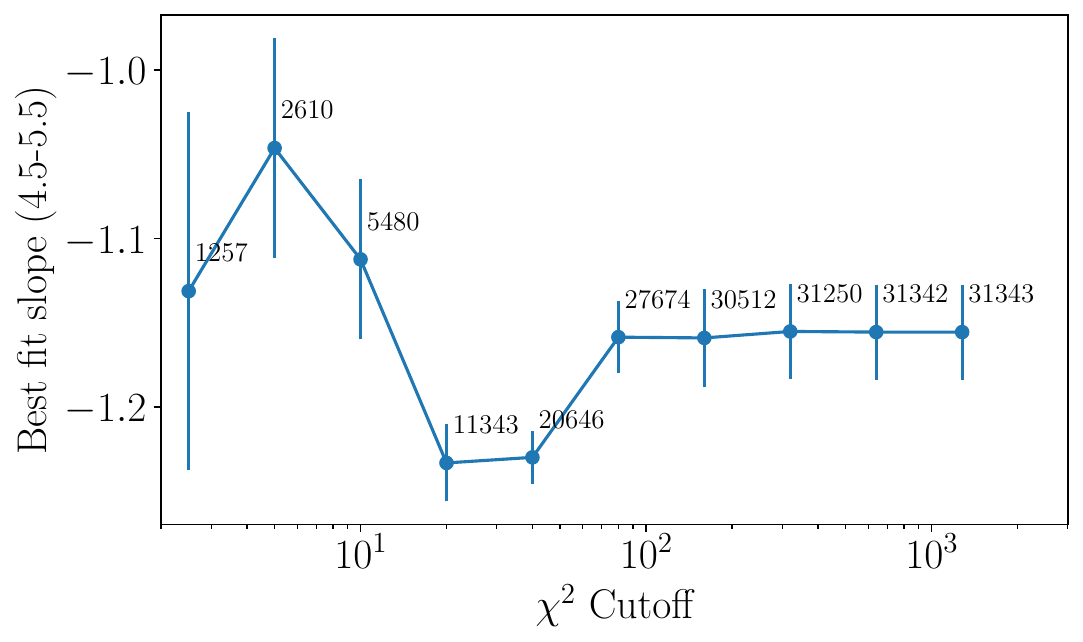}
\caption{Slopes of the linear regression power-law fits with varying cutoffs for the largest permitted minimum $\chi^2$. In all cases, the power-law fit was between log($n_{\rm H_2}$) = 4.5 and 5.5, and only sets with column density $N_{\rm ^{13}CO} \geq 7 \times 10^{15} \textrm{ cm}^{-2}$ were included. Numbers on the plot indicate the number of sets falling within that cutoff, with 31,343 being the total number with the required column density.}\label{fig:chisq_counts}
\end{figure}

We produced additional plots with the same procedure but different values for the largest permitted local minimum $\chi^2$, from 2.5 to 1,280, with the latter limit just sufficient to encompass all of the sets that met the column density requirement $N_{\rm ^{13}CO} \geq 7 \times 10^{15} \textrm{ cm}^{-2}$. All of these plots showed some power-law slope between log($n_{\rm H_2}$) = 4.5 and 5.5, with a value between $\alpha$ = $-1.0$ and $-1.25$ (Figure \ref{fig:chisq_counts}). Uncertainty in the slope is much greater for cutoffs below our main value of $\chi^2 = 10$ but somewhat less for cutoffs above it, with the highest cutoff, including all sets of sufficient column density, having a slope of $\alpha = -1.155 \pm 0.028$.

If there were a strong power-law in the true $\rho$-PDF of Cygnus-X, we might expect the slope of the power-law fit to approach this value more closely for more restrictive cutoffs of $\chi^2$, as we would be sampling from the sets for which our modeling of volume density was most accurate. While there is no clear trend towards a specific value in Figure \ref{fig:chisq_counts}, our power-law fit slopes stay within the range between $\alpha$ = $-1.0$ and $-1.25$, which suggests that the approximate value is still relevant to our conclusions. Running our procedure on synthetic data with less natural noise could help us determine whether there is a systematic bias towards this range of slopes.

\subsection{\tex~Consistency}\label{subsec:consistency}

To confirm the accuracy of our \tex~estimates for CO(2-1) and CO(1-0) for each set that we made in Section \ref{subsec:phys}, we checked them against the \tex~values for \tco(2-1) and \tco(1-0) that RADEX produced alongside its calculated peak radiation temperatures for the best-fit volume density (Step 8). Under LTE conditions, all four excitation temperatures should be equal to each other and to the gas kinetic temperature in the cloud.

\begin{figure}
\epsscale{1.16}
\plotone{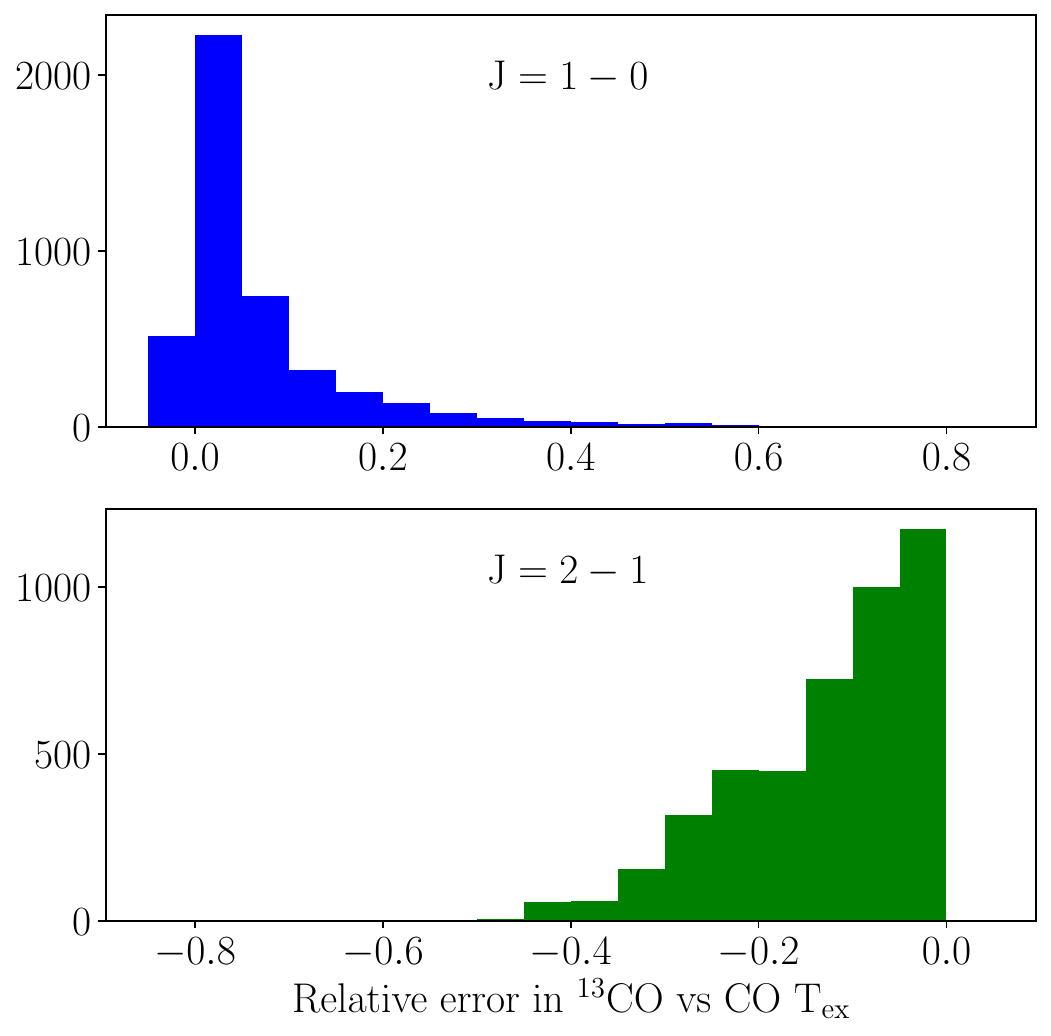}
\caption{Histogram of relative errors between \tex~calculated for CO transition lines and those estimated for the corresponding \tco~lines by RADEX. The horizontal scale goes up to 0.85 to include the highest outlier in the J=1-0 plot (not visible).}\label{fig:errors_tex}
\end{figure}

The results are shown in Figure \ref{fig:errors_tex} for all 4,401 sets with minimum $\chi^2 < 10$ but that did not reach our maximum density of $10^{6}~$\cmvol. (The plot excludes those that reached the highest density since the degeneracy caused an anomalous fit; those sets tend to have error close to zero.) Almost all ${\rm(\sim 87\%)}$ of RADEX's \tco~excitation temperatures for these sets were higher in the J=1-0 line, and all were lower in the J=2-1 line, than our LTE estimates. About 92\% of the sets had errors below 0.3 in both lines, being consistent with our 30\% \tex~error threshold for including them in the first place.

Because we assumed LTE when calculating \tex~in the first place for the CO J=1-0 and J=2-1 transitions, the errors between our computations for CO and RADEX's for \tco~are understandable. RADEX systematically returns lower \tex~for the J=2-1 transition than our calculations because under the non-LTE conditions it simulated, that line is not excited as much, while it usually gives higher \tex~for the J=1-0 line because stimulated emission at high densities contributes to \tex~above what LTE would predict from gas kinetic temperature alone.

\section{Discussion}\label{sec:discuss}

The shallowest $\rho$-PDF slope in 
\citet[][hereafter K11]{2011ApJ...727L..20K} was $-1$, 
which was caused by rotationally supported cores\footnote{Note that the simulations in K11 were also limited by resolutions.}
at a very high density of order $n_{\rm H_2}\sim10^9$ cm$^{-3}$.
Such a density is not observed in our $\rho$-PDF
in Figure \ref{fig:voldens} and could not be identified with these CO lines. In addition, the SMT angular
resolution of $\sim$32\arcsec\ corresponds to a spatial scale
of $\sim$0.22 pc (assuming a distance of 1.4 kpc).
Such a physical scale is not able to resolve the sub-structure
of a dense core. Thus, the shallow slope in our measurement
($-1.12\pm0.05$) is not explained by core rotation.

\citet{2015A&A...577L...6S}, when discussing flattened N-PDFs in the Orion A molecular cloud, suggested that feedback from O-stars in the cloud could be responsible \citep[also see][]{2013ApJ...763...51F}. Similarly, \citet{2013A&A...554A..42R,2014A&A...564A.106T,2015MNRAS.453L..41S,2016MNRAS.461...22P} saw hints of shallow PDFs that were caused by stellar feedback. Cygnus X is a region with its own O-star feedback mainly from the Cygnus OB2 Association, which is to the north of the area we studied. So we tentatively attribute the shallow power-law to the strong feedback environment.

Recently, S22 measured the N-PDF in Cygnus North and Cygnus South regions. Both regions were shown to have two power-law tails in the N-PDF. Cygnus North has a relatively shallow first power-law tail (slope $-1.83$), and its second power-law is even shallower (slope $-1.40$), with the transition being at $A_V\sim80$ mag. The first N-PDF power-law tail for Cygnus South has a slope of $-2.37$ which is the median value for their high-mass star-forming sample, and the second power-law tail is steeper with a slope of $-2.66$, with the transition being at $A_V\sim37$ mag. Our mapped region has more overlap with their Cygnus South region. Due to the limited dynamic range, our fitted $\rho$-PDF (Figure \ref{fig:voldens}) only shows one power-law. So it is not clear if there is a second power-law at higher densities as shown by the N-PDF in S22. If we still follow the K11 scheme, the slope of $-1.12$ for the $\rho$-PDF should correspond to a slope of $-1.19$ for the N-PDF, which is shallower than the measurement from S22. While the comparison between $\rho$-PDF and N-PDF is not straightforward, it seems our measurement indicates a shallower PDF power-law slope than that from S22.

The apparent difference could be due to the different sampling methods (\S\ref{sec:methods}). We are distinguishing different "sets" of aligned velocity components along the same line-of-sight while S22 included the total column in the Herschel map. We are only including sets in LTE ($T_{\rm ex}$ agrees within 30\% for two $^{12}$CO transitions) which intrinsically prefers denser gas. We are keeping preferentially higher column densities to ensure we are in the power-law regime. Considering these limits, we are effectively sampling deeply embedded denser gas that is temporarily shielded from the external radiation. But this gas probably feels the compression due to the high pressure from hot gas outside. This special sampling probably leads to the relatively shallower $\rho$-PDF. In the future, a comparison with synthetic observations of gas that is strongly impacted by feedback could help distinguish between such processes.

Our finding of a power-law $\rho$-PDF generally agrees with S22. However, \citet{2023MNRAS.523.1373M} did not find power-law profiles in $^{13}$CO N-PDF in the majority of their sample. Note that they broke the Cygnus-X complex into multiple individual clouds in the position-position-velocity space. While \citet{2023MNRAS.523.1373M} considered the optical depth effect for the J=1-0 transition and argued that it was a minor effect, they did not clearly show how they address the depletion effect of CO. Since they assumed LTE with median $T_{\rm ex}\sim$10-16K, the depletion of $^{13}$CO could impact the column density derivation and thus the N-PDF, assuming the dust temperature is similar to $T_{\rm ex}$ under LTE. For instance, \citet{2021ApJ...908...76L} showed noticeable depletion when dust temperature was below 20 K, with the depletion factor increasing from 2 to 10 at a temperature of 12 K. If higher column density regions have lower temperature, the correction for depletion should be larger at high column density. Indeed, \citet{2016A&A...587A..74S} found that CO depletion probably caused the N-PDF falloff at $A_V\gtrsim40$ mag. They concluded that N-PDFs based on molecular lines were not well constrained.

\section{Summary and Conclusion}\label{sec:conclude}

We have presented OTF maps of the Cygnus-X molecular cloud complex in the J=2-1 transition of both CO and $^{13}$CO using the ARO SMT telescope. The maps extend from 20$^{\rm h}$24$^{\rm m}$~to~20$^{\rm h}$39$^{\rm m}$~in RA and from 38\arcdeg~to~40\arcdeg43\arcmin~in DEC, covering an area of $\sim$10 square degrees. The mapped area overlaps with the Nobeyama 45m Cygnus-X CO Survey in the southern portion of the complex in which we carry out a multi-transition analysis for CO(1-0), $^{13}$CO(1-0), CO(2-1), and $^{13}$CO(2-1). Through the analysis, we have derived physical properties of the molecular gas, especially the gas volume density. 

We use {\it astrodendro} to identify velocity-coherent gas components (sets in each pixel) in all four lines. For each set we fit a Gaussian profile to each line spectrum and obtain the peak intensity and line dispersion. We derive the excitation temperature $T_{\rm ex}$ for each CO line assuming optically thick conditions. We select sets whose $T_{\rm ex}$ agree within 30\% between the two CO lines so that LTE is a reasonable assumption for the sets. We then derive the $^{13}$CO column density for each set. These physical properties are fed to the RADEX program with which we iteratively find the best H$_2$ volume density by matching the intensities between the model and the observation for the two $^{13}$CO lines. We construct the volume density probability distribution function ($\rho$-PDF) which shows a single power-law tail between log($n_{\rm H_2}$)= 4.5 and 5.5 with a slope of $\sim -1.12\pm0.05$. Higher transition lines are needed to trace higher densities.

The $\rho$-PDF power-law slope is shallower than the theoretical values between $-1.5$ and $-2.0$ in the context of self-similar gravitational collapse \citep[e.g.,][]{2011ApJ...727L..20K}. Within the same context, the corresponding column density PDF power-law tail is also shallower than the measurements from \citet{2022A&A...666A.165S} using Herschel maps. Possible causes include the fact that Cygnus-X is strongly impacted by feedback from the OB2 association \citep[similar to what was found in][]{2015A&A...577L...6S} and we are specifically sampling deeply embedded dense gas. Future comparison with synthetic observations with feedback may help clarify these causes. Our observational study of $\rho$-PDF provides another useful test for theories.

\acknowledgments

We thank the anonymous referee for a constructive report. We acknowledge fruitful discussions with Yancy Shirley and Samantha Scibelli. The Heinrich Hertz Submillimeter Telescope is operated by the Arizona Radio Observatory, which is part of Steward Observatory at The University of Arizona. We thank Patrick Fimbres, Christian Holmstedt, Robert Moulton, Blythe Guvenen, and all the observatory staff members for helping with the data acquisition and maintaining a smooth telescope operation. 
\newline
\software{Python \citep{python}, SciPy \citep{scipy}, Astropy \citep{Astropy-Collaboration13}, Numpy \citep{numpy}, Matplotlib \citep{matplotlib}, SAOImageDS9 \citep{2003ASPC..295..489J}, RADEX \citep{2007A&A...468..627V}, spectral-cube \citep{adam_ginsburg_2014_11485}}

\facility{SMT,NRO45}

\bibliography{ref}
\bibliographystyle{aasjournal}

\end{document}